\documentclass[11pt]{article}
\usepackage{supertech}

\notesfalse

\title{\vspace*{.09in}\bf
       {\Large Scheduling Algorithms for Procrastinators}}

\author{
        Michael A. Bender\thanks{Department of Computer Science,
            Stony Brook University,
            Stony Brook, NY 11794-4400, USA.
            Email:~\texttt{bender@cs.sunysb.edu}.
            This research was supported in part by NSF Grants
            CCR-0208670,
            CCF-0621439/0621425,         
            CCF-0540897/05414009,        
            CCF-0634793/0632838,         
            and CNS-0627645.             
}
    \and
        Rapha\"{e}l Clifford\thanks{Department of Computer Science,
            University of Bristol,
            Merchant Venturers Building,
            Woodland Road,
            Bristol BS8 1UB, UK.
            Email:~\texttt{clifford@cs.bris.ac.uk}.}
    \and
         Kostas Tsichlas\thanks{Computer Engineering and Informatics Department,
            University of Patras,
            26500 Patras, Greece.
            Email:~\texttt{tsihlas@ceid.upatras.gr}.}
}

\date{}

\newcommand{\lad}{\lrtb}
\newcommand{\lrtb}{LRTB\xspace}
\newcommand{\opt}{OPT\xspace}
\newcommand{\srpt}{SRPT\xspace}
\newcommand{\edd}{EDD\xspace}
\newcommand{\lssf}{LSSF\xspace}
\newcommand{\fifo}{FIFO\xspace}
\newcommand{\thrashing}{\textsc{Thrashing}\xspace}
\newcommand{\delay}{$\alpha$-DLY\xspace}
\newcommand{\extended}[1]{\tilde{d}_{#1}}
\newcommand{\extendedj}{\extended{j}}

\begin{document}

\maketitle

\begin{quote}\begin{quote}\footnotesize
  If once a man indulges himself in murder, very soon he comes to think little
  of robbing; and from robbing he comes next to drinking and
  Sabbath-breaking, and from that to incivility and procrastination.

  \vspace{0.6ex}

  --- Thomas de Quincey

   \vspace{1.ex}

\end{quote}\end{quote}

\begin{abstract}
This paper presents scheduling algorithms for procrastinators, where
the speed that a procrastinator executes a job increases as the due
date approaches.  We give optimal off-line scheduling policies for
linearly increasing speed functions.  We then
explain the computational/numerical issues involved in implementing
this policy. We next explore the online setting, showing that there
exist adversaries that force any online scheduling policy to miss due
dates. This impossibility result motivates the problem of minimizing
the \emph{maximum interval stretch} of any job; the interval stretch
of a job is the job's flow time divided by the job's due date minus
release time. We show that several common scheduling strategies,
including the ``hit-the-highest-nail'' strategy beloved by
procrastinators, have arbitrarily large maximum interval stretch.
Then we give the ``thrashing'' scheduling policy and show that
it is a $\Theta(1)$ approximation algorithm for the maximum interval
stretch.
\end{abstract}

\secput{intro}{Introduction}
We are writing this sentence two days before the deadline.
Unfortunately that sentence (and this one) are among the first that
we have written. How could we have delayed so much when we have
known about this deadline for months?  The purpose of this paper is
to explain why we have waited until the last moment to write this
paper.

In our explanation we model procrastination as a scheduling problem.
We cannot use traditional scheduling algorithms to model our
behavior because such algorithms do not take into account our (and
humanity's) tendency to procrastinate. The advantages of
procrastination are well documented: the closer to a deadline a task
is executed, the less processing time the task appears to require.
Hence, it is common for a person to delay executing some onerous job
in order to spend as little time as possible working on it.

Regarding this paper, it will certainly be written quickly --- it
will have to be, since the deadline is near.  Perhaps we will write
faster under pressure because we will expend less time overanalyzing
each design option. Other aspects of the paper may change because of
this time pressure. In any case, the writing will proceed faster
than if we had begun earlier.

Our scheduling problem for procrastinators is unusual in that the
processing time of a job depends on the times when the job is run.
We are given as input a set of jobs $\mathcal{J}=\{1,2,\ldots,n\}$.
Each job $j$ has release time $r_j$, due date $d_j$, and work $w_j$;
without loss of generality, we assume that the jobs are indexed by
increasing release times. \emph{Preemption} is allowed; that is, a
running job can be interrupted and resumed later. The speed at which
job $j$ is run depends on the times that $j$ is executed; the closer
to the due date $d_j$, the faster $j$ can be executed. Specifically,
\emph{speed function} $f_j(t)$ indicates that at time $t$, job $j$
is executed with speed $f_j(t)$; thus, if $j$ is executed during
time interval $[a,b]$, then $\int_{t =a}^{b}f_j(t)\,dt$ units of
work of job $j$ complete.

Throughout most of the paper we focus on \emph{linear} speed
functions.
We assume that when job $j$ first is released, it is executed with
speed $0$. In accordance with this last assumption, when the call
for papers first appeared, we snapped into action and accomplished
nothing.

Despite our whimsical and self-referential style, we hope to
emphasize that the scheduling problems on job streams with
time-dependent processing times have mathematical subtlety as well
as practical relevance. The time-dependent processing models in this
paper may be useful for industry and sociology because they give
better scheduling models of human behavior; no model can truly be
accurate that does not account for people's ability to work faster
under the temporary stress of deadlines. More generally, many common
scheduling problems in both daily life and industry have tasks whose
processing times are time-dependent. For example, an airplane that
is late in arriving may have the boarding procedure expedited, a
construction project that is behind may have more workers assigned
to it, and a shipment that is late may be delivered faster by using
an alternative, more expensive means of transportation. Indeed a
major reason for the success of companies such as Fedex, UPS, and
DHL is that the world is filled with scheduling problems executed by
procrastinators.

\subsubsection*{Related Work}

A number of other optimization problems have well studied
time-dependent variants, including work on time-dependent shortest
paths~\cite{OrdaRom90} and time-dependent
flows~\cite{FleischerSkutella03,FleischerSkutella02}.
Some authors, typically in the operations-research community, have
also worked on scheduling with time-dependent processing times (see,
e.g.,~\cite{AlidaeeWomer1999,BachmanJaniakKovalyov02,GawiejnowiczKurcPankowska02,GawiejnowiczPankowska95b}),
but for the offline and nonpremptive case.
Of course, preemptive and online models are best for modeling the
behavior of procrastinators, who tend to timeshare and thrash as the
deadlines approach. Moreover, our introduction of preemptive
scheduling with time-dependent processing times requires an entirely
different model.  Previous work has assumed that the processing time
$p_j(t)$ for job $j$ is a function of the starting time $t$. We
cannot have such a model in a preemptive case because the job may be
executed during many different time intervals.  This issue motivates
our need for processor speeds: job $j$ is executed with speed
$f_j(t)$ at time $t$; the processing time is the sum over all
intervals when job $j$ is executed, and the integral of $f_j(t)$
over all times that the job is executed must equal the job's work.
Curiously, if we analyze existing nonpreemptive models (e.g.,
linearly decreasing processing times) and analyze what processor
speeds and total work must be to generate these processing times,
then we can create instances where the processing speeds approach
infinity; clearly such a model is unrealistic.

The most closely related work in the literature is on scheduling
algorithms for minimizing power consumption and, in particular, on
``speed scaling.''
See~\cite{BansalKiPr04,PruhsStUt05,Bunde07,BansalPr05,UysalPrGa02,AlbersFu06}
for some recent results and~\cite{IraniPr05} for an excellent
survey.  The idea of speed scaling is that the processing speed of a
job is variable, but faster speeds consume more power. This ability
to vary the speeds is reminiscent of the procrastinator who can run
at unsustainable rates near the deadline.   However, unlike in the
speed-scaling model, the procrastinator has less freedom in choosing
the processing speed; the processing speed is solely determined by
the proximity to the deadline.

We note that there exist other scheduling papers where processors
have different speeds, both for ``related''
processors~\cite{ChudakShmoys99,ChekuriBender01,BenderRabin02} and
for ``unrelated''
processors~\cite{DavisJaffe81,LawlerLabetoulle78,JansenPorkolab99}.
However, neither situation models procrastination scheduling (or
speed scaling), where the processing speeds per job change over
time.

There are other scheduling problems on how to schedule reluctant
workers, such as the lazy bureaucrat
problem~\cite{ArkinBenderMitchellSkiena99,HepnerStein02,%
ArkinBenderMitchellSkiena03}. However, the lazy bureaucrats in the
scheduling problem are trying to accomplish as few of the jobs as
possible, whereas the procrastinators in the current scheduling
problem are trying to finish all of the jobs.

\subsection*{Results}
In this paper we present the following results.
\begin{itemize}

\item \emph{Optimal offline scheduling ---}  We first give \emph{optimal}
offline scheduling policies for the case where a scheduling instance
has a feasible solution. We consider the case of linear speed
functions, $f_j(t)=m_j(t-r_{j})$, for constant $m_j\geq0$. (In the
offline problem, the scheduler sees the entire problem instance
before it has to begin scheduling.) Specifically, the policy gives
the feasible solution in which the processors spend the minimum
total time running. These results are consistent with a
procrastinator who, after missing crucial deadlines, muses ``if I
could do it all over again\ldots.''

\item  \emph{Computational/numerical issues ---} We show that, curiously,
despite a simple optimal scheduling policy, actually
\emph{determining feasibility} of the resulting schedule is not even
known to be in NP. In particular, determining feasibility is hard
because of the computational difficulties of summing square roots.
We know of few scheduling problems where this
intriguing issue arises.

\item
\emph{Online scheduling ---}  We next turn to  online scheduling.
Not surprisingly, the feasibility problem is not achievable in an
online setting.  In particular, even if the online procrastinator
has a feasible set of jobs, he/she may be forced to miss an
arbitrarily large number of due dates.

\item
\emph{Online maximum interval stretch ---} A procrastinator may be
forced to execute jobs beyond their due dates, that is, for some job
$j$, the completion time $C_j$ may exceed the due date $d_j$.
Generally speaking, if a procrastinator has a year to do a job $j$,
and completes $j$ two weeks late, the situation is better than if
the procrastinator has only one day to do $j$, but completes two
weeks late.
This observation motivates the notion of \emph{interval stretch},
defined as the flow time (time the job spends in the system) divided
by the job's interval.  More formally, the interval
stretch\footnote{This definition deviates from the standard notion
of stretch where the flow time is divided by the total time the job
has spent working~\cite{BenderChMu98}. However, it is appropriate
here as jobs have due dates which can be missed and job speed is
time-dependent.} of job $j$ is defined as $s_j=(C_j-r_j)/(d_j-r_j)$.
We consider the optimization metric \emph{maximum interval stretch}
(abbreviated to \emph{max-stretch}), $\max_{j}s_j$.

We study online scheduling of feasible scheduling instances. We
explore traditional scheduling policies for the procrastinator, such
as First-In-First-Out (FIFO), Shortest-Remaining-Processing-Time
(\srpt), and earliest-due-date (\edd). We show, not surprisingly,
that these policies do not perform well and can lead to unbounded
max-stretch.
A common scheduling policy among
many procrastinators is ``hit-the-highest-nail'', that is, execute
the task that most crucially requires attention, formally,
Largest-Stretch-So-Far (\lssf).  In \lssf we execute the job in the
system that \emph{currently} has the largest interval stretch. We
prove, perhaps surprisingly, that \lssf can lead to arbitrarily
large max-stretch.
We conclude our exploration of max-stretch by exhibiting an online
algorithm for the procrastinator, \thrashing, that yields
$\Theta(1)$ max-stretch.
This last result holds even when each job has a bound on its maximum
execution speed.

\end{itemize}

\secput{offline}{Offline Procrastination Scheduling}

In this section we consider the \emph{offline}
procrastination-scheduling problem. First, we give an optimal
scheduling policy based on a simple priority rule.   Then we show
that it is computationally difficult to determine whether a
scheduling instance is feasible, despite this priority rule. We
focus on linear speed functions, $f_j(t)=m_j(t-r_{j})$. We will show
that, without loss of generality, we can assume that all speed
functions can have unit slope, i.e., that $m_j=1$.

\subsection*{Optimal Offline Scheduling Policy}

We now give an optimal scheduling policy for the offline
procrastination problem based on a simple priority rule.

We first define terms.  We say that a schedule is \emph{feasible} if
all jobs complete within their intervals; we say that a feasible
schedule is \emph{optimal} if the total processing time is
minimized. Observe that if an optimal schedule has no idle time
then all feasible schedules are also optimal.

The optimal algorithm starts at the latest due date and works
backwards in time, prioritizing jobs by the latest release time.
Whenever a new job is encountered (at the job's due date) or a job
completes, then the job in the system having the latest release time
is serviced. Where two or more jobs have the same release time the
scheduler chooses between them in an arbitrary but fixed way.  We
call this scheduling algorithm \emph{Latest Release Time Backwards
(\lrtb)}.

Observe that \lrtb is the traditional Earliest Due Date (\edd)
policy (see, e.g.,~\cite{KargerStWe98}) when we reverse the flow of
time so that release dates become due dates and due dates become
release dates.  In traditional scheduling, time can flow in either
direction, so that both \lrtb and \edd generate feasible schedules.
In contrast, in the procrastination problem, \edd performs poorly;
see \secref{online}. The intuition of the algorithm is that it
always tries to push the work of a job as near to its due date as
possible in order to maximize the processing speed.

Observe that the job priorities depend only on the release times and
not the slopes. This lack of dependence on the slopes should not be
surprising because we can transform any scheduling instance into an
instance having all unit slopes by rescaling each job $j$'s work to
be $w'_j=w_j/m_j$. Alternatively, we could give all jobs unit
maximum speeds, $f_j(d_j)=1$, by setting $m_j=1/(d_j-r_j)$ and then
rescaling the work. Consequently, in the rest of the paper, we
assume that the job slopes are $1$, unless otherwise stated.

\begin{figure}[t]
\protect\figlabel{thmone-figure} {\centering
 \includegraphics[scale=0.7]{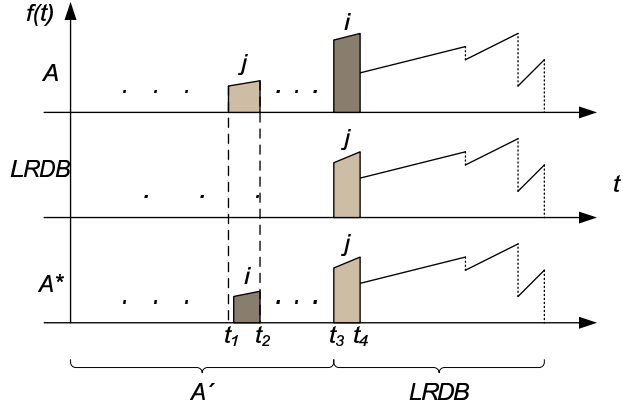}
 \caption{Schedule $A^*$ results from the merge of $A'$ and \lrtb.
 Schedule $A'$ results from $A$ by exchanging jobs $j$ and $i$.
 The small gap after $t_{1}$ indicates that this exchange is more time
efficient.}}
\end{figure}

In the following we prove that algorithm \lrtb produces the optimal
schedule.

\begin{theorem}\thmlabel{same_constant}
\lrtb is an optimal algorithm for the procrastination scheduling
problem. There is a unique optimal solution provided that the
release time of the jobs are distinct.
\end{theorem}

\begin{proof}
The proof is by an exchange argument. We first assume that no two
jobs have the same release time and then relax that assumption at
the end. Suppose for the sake of contradiction that there exists an
optimal schedule $A$ different from \lad. Specifically, these
schedules differ in the order of execution of two jobs with
different release times. We perform a single exchange of work to
yield another feasible schedule $A^*$ having smaller total
processing time than $A$, thus obtaining a contradiction.

Consider the latest instant in time where \lad
differs from $A$ and call this time $t_4$. Consider an arbitrarily
small interval $[t_3,t_4]$, when job $j$ runs in \lad and job $i$
runs in $A$. See
Figure~1 for a depiction of the setting. By the definition of \lad,
$i$, $j$, and $t_4$, $r_i<r_j$. Consider some earlier time interval
$[t_1,t_2]$, i.e., $t_2\leq t_3$, when job $j$ runs in $A$. Define
$t_1$, $t_2$, and $t_3$ so that the amount of work that can be
executed on job $j$ is the same, that is,
$$
\int_{t=t_1}^{t_2}f_j(t)\,dt=\int_{t=t_3}^{t_4}f_j(t)\,dt.
$$

Now we make a new schedule $A^*$ from $A$ by exchanging the work
done during intervals $[t_1,t_2]$ to $[t_3,t_4]$. Specifically in
$A^*$, job $j$ is run during $[t_3,t_4]$ and job $i$ is run during
$[t_1,t_2]$. We know that this exchange is allowed because $d_j>t_4$
(from the \lad and $A$ schedules) and because $r_i<r_j\leq t_1$
(from the  $A$ schedule and because $r_i<r_j$).  By the definition
of the intervals, the same amount of work on $j$ can be done during
each interval. Computing the area of the trapezoids defined by
$f_j(t)$, we obtain
$$
\left(t_4-t_3\right)\left(\frac{t_4+t_3}{2}-r_j\right)m_j
=\left(t_2-t_1\right)\left(\frac{t_2+t_1}{2}-r_j\right)m_j,
$$
meaning that
\begin{equation}\eqlabel{one}
(t_4^2-t_3^2)/2-r_j(t_4-t_3)=(t_2^2-t_1^2)/2-r_j(t_2-t_1).
\end{equation}
Observe that $t_4-t_3<t_2-t_1$ because the speed that $j$ is
executed during $[t_3,t_4]$ is greater than during $[t_1,t_2]$.

The amount of work on job $i$ that needs to be exchanged from
$[t_3,t_4]$ to $[t_1,t_2]$ is $(t_4^2-t_3^2)/2-(t_4-t_3)r_i$. But
since $r_i<r_j$ and $t_4-t_3<t_2-t_1$,
\begin{equation}\eqlabel{one-a}
(r_j-r_i)(t_4-t_3)<(r_j-r_i)(t_2-t_1)
\end{equation}
From \eqreftwo{one}{one-a}, we obtain the inequality
$$
(t_4^2-t_3^2)/2-r_i(t_4-t_3)<(t_2^2-t_1^2)/2-r_i(t_2-t_1),
$$
and therefore interval $[t_1,t_2]$ is big enough to execute all of
the work on job $i$ and still leave some idle time. Hence, schedule
$A^*$ is feasible and spends a smaller amount of time working. This
gives us our contradiction.

We now explain the case where two jobs $1$ and $2$ have the same
release time. Assume that job $1$ is scheduled to execute some work
in the time interval $[t_1,t_2]$ and job $2$ is scheduled to execute
some work in the interval $[t_3, t_4]$.  If we exchange the work for
jobs $1$ and $2$, the relationship between the new time intervals
and the old is expressed by the simple equation $t_4^2-t_3^2 =
t_2^2-t_1^2$. Therefore the total time to execute both jobs remains
the same after exchange.  As a result, the order in which these jobs
are executed does not affect the total processing time, and so \lrtb
is an optimal algorithm no matter what the tie-breaking rule is.
This completes the proof. \qed
\end{proof}

\subsection*{Determining Feasibility May Not Be in NP}

One of the remarkable features of the procrastination
problem is that, despite having the simple optimal scheduling policy
\lrtb, it is unclear whether determining the feasibility of a
scheduling instance is even in NP, even for linear speed functions.

The difficulty is numerical. Calculating the actual processing time
of the job $j$ given a starting or ending time $t$ and speed
function $f_j(t)=t-r_j$ requires computing square roots. Determining
the feasibility of the schedule therefore requires computing sums of
square roots and their relationship to an integer, and this problem
appears to be numerically difficult.

The basic sum-of-square-roots problem is to determine whether
 $$
 \sum_{i=1}^{m}{\sqrt{x_i}} \geq I
 $$
for some $x_i,I\in\mathds{Z}\,\, (1\leq i \leq m)$. Because there is
no known polynomial-time algorithm for deciding the
sum-of-square-roots problem,
basic computational-geometry problems such as Euclidean TSP or
Euclidean shortest paths are not known to be in NP. See the Open
Problems Project~\cite[Problem 33]{DemaineMitchellORourke-topp}
(originally from~\cite{ORourke81}) and the Geometry
Junkyard~\cite{Eppstein07} for nice discussions of the
sum-of-square-roots problem.

We establish the difficulty of procrastination scheduling by providing
a reduction from any instance of the sum-of-square-roots problem. To derive the
cleanest reduction, we allow the existence of nonlazy jobs, i.e.,
jobs that are always executed at the same speed, i.e., having slope
$0$. (It is likely that a reduction can be made to work using no
nonlazy jobs, but at the cost of additional complications.)

\begin{theorem}
The procrastination scheduling problem is not decidable in
polynomial time unless the sum-of-square-roots problem is decidable
in polynomial time. The procrastination scheduling problem is not in
NP unless the sum-of-square-roots problem is also in NP.
\end{theorem}

\begin{proof}
We reduce the sum-of-square-roots problem to the procrastination
scheduling problem. Given integers $x_1,\ldots,x_{n-1}$ and $I$,
we will create a procrastination-scheduling problem with $n$ jobs.
The procrastination scheduling problem will be feasible if and only
if  $\sum_{i=1}^{n-1}{\sqrt{x_i}} \geq I$.

We first give the structure of the scheduling instance and then
determine the release times, deadlines, and work for each job. In
our scheduling instance, jobs $1\ldots n-1$ have
nonoverlapping intervals, so that $r_1=0$, and the due date of one
job is the release date of the next: $r_{i+1}=d_i$
($i=1,\ldots,n-2$).
The speed functions have slope~$1$.
Job $n$ is nonlazy. We place this job's interval so that it
overlaps with the intervals of all other jobs, i.e.,
$r_{n}=r_1$ and $d_{n}=d_{n-1}$.

We now specify jobs $1,\ldots,n-1$. For job $i$, we choose
interval length $\ell_{i}$ ($=d_i-r_i$) and work $w_{i}$ to be
positive integers such that $\ell_{i}^{2}-2w_{i}=x_{i}$; many
choices of $\ell_{i}$ and $w_i$ will work. It suffices to choose
positive integers $\ell_i$ and $w_i$ such that
$0<\ell_i^2-2w_i<\ell_i$. For example, by choosing $\ell_i=x_i+2$
and $w_i=({x_i^2+3x_i+4})/{2}$, all conditions are fulfilled. Note
that $x_i^2+3x_i+4$ is always an even number for $x_i>0$ and thus
$w_i$ is an integer.

Each job $i$ ($i=1,\ldots,n-1$) runs fastest
when pushed to the right side of its interval.
We show that such a job runs in time
$t_{i}=\ell_{i}-\sqrt{\ell_{i}^2-2w_{i}}$.
To establish this running time, we set up and solve a quadratic
equation.
By simple geometry, we have the following relationship
between running time $t_{i}$ and work $w_{i}$:
$$
w_{i}=t_{i}\left(\ell_{i}-{t_{i}}/{2}\right).
$$
This quadratic equation has two roots,
$$
t_{i}=\ell_{i}\pm\sqrt{\ell_{i}^{2}-2w_{i}},
$$
and the smaller root is the running time of the job.
(This can be seen since the larger root is greater than $\ell$,
the interval length.)

The total time taken by all $n-1$
nonoverlapping jobs when scheduled optimally is therefore
 $$
    \sum_{i=1}^{n-1}\ell_i-\sum_{i=1}^{n-1}{\sqrt{\ell_i^2-2w_i}}
 =  \sum_{i=1}^{n-1}\ell_i-\sum_{i=1}^{n-1}{\sqrt{x_i}}.
 $$

We now construct the nonlazy job $n$. As described earlier
$r_{n}=0$ and $d_{n}=d_{n-1}$. We set work $w_{n}=I$.

There is a feasible solution for this scheduling problem
if and only if
$$
d_{n}\geq
w_{n}+\sum_{i=1}^{n-1}\ell_i-\sum_{i=1}^{n-1}{\sqrt{x_i}}.$$
This is
the case, as long as $I\leq\sum_{i=1}^{n-1}{\sqrt{x_i}}$, since by
construction, $d_{n}=d_{n-1}=\sum_{i=1}^{n-1}\ell_i$.
Thus, an
arbitrary instance of the sum-of-square-roots problem can be reduced
to an instance of procrastination scheduling, implying the numerical
difficulty of procrastination scheduling.\qed\end{proof}

\secput{online}{Online Algorithms}

This section considers the online procrastination scheduling
problem. In the online problem, jobs $1\ldots n$ arrive over time.
Job $j$ is known to the scheduler only at the release time $r_j$, at
which point the scheduler also learns the values of $w_j$ and $d_j$.
We first show that it is difficult for an online scheduler to find
feasible schedules. Next we search for online algorithms that
generate small, ideally constant, max-stretch. We show that
traditional scheduling policies such as \edd, \srpt, and \fifo, have
large, typically unbounded, max-stretch. We next consider the
scheduling policy Largest-Stretch-So-Far (\lssf), which executes the
job in the system currently having the largest interval stretch.
This policy formalizes the ``hit-the-highest-nail'' scheduling
policy, that is, execute the task in the system that most crucially
requires attention.
 More precisely, in the \lssf scheduling
policy, we run the job in the system that has incurred the largest
interval stretch so far, that is, at time $t$ we execute the job $j$
that maximizes $(t-r_j)/(d_j-r_j)$.  We show that, remarkably, \lssf
also has unbounded max-stretch. We conclude this section by
exhibiting the scheduling algorithm \thrashing, whose max-stretch is
within a constant factor of optimal and then give a generalization
to non-linear speed functions.
One consequence of this last result is that good online
max-interval-stretch bounds are achievable even when the
procrastinator's maximum processing speed is at most a constant
factor faster than a nonprocrastinator's speed.

\subsection*{Basic Results}

We first show that any online algorithm can be forced to miss due
dates, even when the scheduling instance is feasible. A job~$j$ has
\emph{slack} if the work, $w_j$, associated with it is less than the
area between $r_j$ and $d_j$, i.e., $w_j <(d_j-r_j)^2/2$.

\begin{figure}[t!]
\centering
{\sf (a)}~~\includegraphics[scale=0.8]{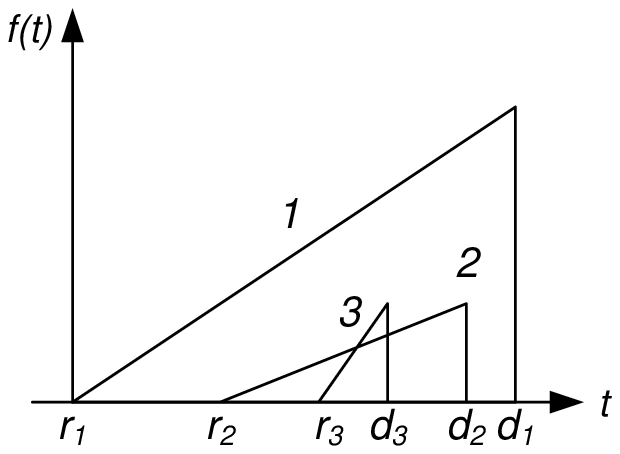}~~~~~~~~~~%
{\sf (b)}~~\includegraphics[scale=0.8]{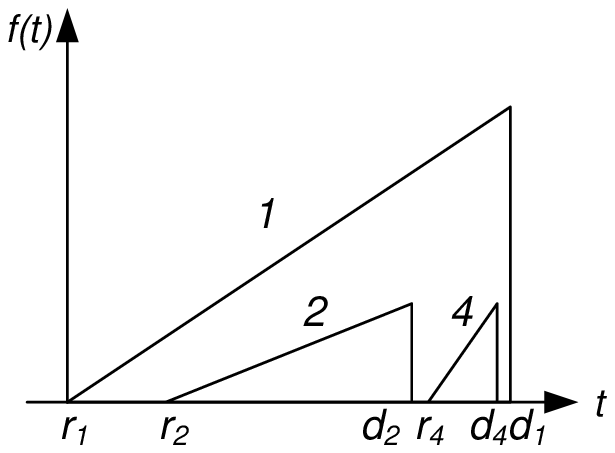}
\caption{(a)~Case~1: job~$1$ is run at time $r_2=3$.  Then job~$3$
arrives. Either job~$2$ or~$3$ is forced to miss its due date.
(b)~Case~2: job~$2$ is run at time $r_2=2$.  Then job~$4$ arrives.
Either job~$1$,~$2$, or~$4$ is forced to miss its due
date.\figlabel{hard1and2}}
\end{figure}

\begin{theorem}\thmlabel{feasible}
For any online algorithm, there is a feasible job stream
on which that algorithm misses due dates.
\end{theorem}

\begin{proof}
We show that regardless of the online scheduling decisions, the
adversary can force the algorithm to miss due dates by maliciously
selecting future jobs. The adversary first sends jobs~$1$ and~$2$,
where $r_1<r_2$ and $d_2<d_1$. Both jobs~$1$ and~$2$ have some slack
and the set $\{1,2\}$ is feasible. At time $r_2$ there are two
cases:

\begin{enumerate}
\item Job~$1$ is serviced at time $r_2$.  Then the adversary
places a job~$3$ with $r_2<r_3<d_3<d_2$. Job~$3$ is designed so that
the entire interval $[r_2,d_2]$ is required to complete jobs~$2$
and~$3$ by their due dates.  Since the online algorithm works
partially on job~$1$ during this interval, either job~$2$ or~$3$
misses its due date; see \figref{hard1and2}(a).

\item Job~$2$ is serviced at time~$r_2$.  The adversary
places a job~$4$ with $r_4>d_2$ and $d_4<d_1$.  Job~$4$ is designed
so that all the time between~$r_2$ and~$d_1$ is required to complete
jobs $1$, $2$, and $4$ by their due dates.  However, as job $2$ has
some slack we know that by \thmref{same_constant} that the optimal
strategy is to run $1$ at time $r_2$ and that this strategy is
unique. Therefore, by running $2$ at time $r_2$ the algorithm misses
at least one of the due dates; see
\figref{hard1and2}(b).\qed
\end{enumerate}
\end{proof}

Observe that, as stated, this example has job parameters that may be
irrational (because of square roots). In fact, we can round job
parameters so that all are rational and the input size (number of
bits) is polynomial in $n$.

By repeating this construction, the adversary can force the
algorithm to miss an arbitrarily large number of due dates. Thus,
\thmref{feasible} explains why procrastinators may have a harder
time juggling online tasks than non-procrastinators.

We now show that most traditional scheduling policies for
non-procrastinators do not work well for procrastinators. The
following theorem gives the performance of First-In-First-Out (\fifo),
Earliest-Due-Date (\edd), and Shortest-Remaining-Processing-Time
(\srpt).

\begin{theorem}
There exist feasible scheduling instances of a constant number of jobs for
which the max-stretch of the First-In-First-Out (\fifo) and
Earliest-Due-Date (\edd) scheduling policies can be arbitrarily
large.
There exist feasible scheduling instances of $n$ jobs for which the
Shortest-Remaining-Processing-Time (\srpt) scheduling policy achieves a max-stretch of
 $\Theta(\sqrt{n})$.
\end{theorem}

\begin{proof}
There is a bad example for \fifo consisting of only two jobs. Let
$r_1<r_2<d_2<d_1$. Set $w_1$ and $w_2$ so that optimal schedule is
to execute job~$2$ to completion as soon as it arrives, and then
finish job~$1$. In \fifo, job~$2$ will not start work until job $1$
has completed and will finish late.  The interval stretch of job~$2$
can be made arbitrarily large by decreasing $w_2$ and $d_2-r_2$ or
by increasing $d_1$ and $w_1$.

There is a bad example for \edd consisting of only three jobs. As
before, let $r_1<r_2<d_2<d_1$. In \edd, job~$2$ is executed starting
at its arrival time $r_2$ because this job has the earliest
deadline. By the proof of \thmref{same_constant}, job $1$ can be
made finish its work after its due date.  Now set a third job so
that $r_3=d_1$ and $d_3-r_3$ is small compared to the lateness of
job~$1$.  The interval stretch of job~$3$ can be made arbitrarily
large by decreasing $d_3-r_3$ or by increasing the lateness of
job~$1$.

There is a bad example for \srpt consisting of $n$ jobs. All jobs
are released at time $0$. Give job~$1$ the largest amount of work:
$w_1=1$. Give all other jobs $w_2=w_3=\cdots=w_n=1/2$. Set $d_1$ so
that job~$1$ must be executed as soon as it arrives in order not to
be late, i.e., $d_1=2$. Give all other jobs later deadlines:
$d_2=d_3=\cdots=d_n=\sqrt{n}+2$. In the optimal schedule, job~$1$ is
executed first and the remaining jobs are executed in any order. In
contrast, in \srpt, jobs~$2\ldots n$ are executed before job~$1$.
One job will be completed at time $1$, the next at time $\sqrt{2}$,
the next at time $\sqrt{3}$, and the last at time $\sqrt{n-1}$. (A
calculation similar to this is explained in greater detail in the
next section.) Only after all other jobs complete does job~$1$
complete, giving it an interval stretch of $\Omega(\sqrt{n})$.\qed
\end{proof}

\subsection*{Hitting the Highest Nail Does Not Work}

A common scheduling strategy among procrastinators is
``hit-the-highest-nail,'' that is, execute the job that is farthest
behind. Since the objective is to minimize the max-stretch,
``hitting-the-highest-nail'' translates to running the job that has
the largest interval stretch.  We call this strategy
\emph{Largest-Stretch-So-Far (\lssf)}.  More precisely, in the \lssf
scheduling policy, we run the job in the system that has incurred
the largest interval stretch so far, that is, at time $t$ we execute
the job $j$ that maximizes $(t-r_j)/(d_j-r_j)$. Thus, the algorithm
might execute a job $i$, but switch to a smaller job $j$ that
arrived after $i$, once $j$'s interval-stretch-so-far surpasses that
of $i$'s.

Remarkably, even for feasible scheduling instances, \lssf may
schedule jobs to have unbounded max-stretch. Below we exhibit such
an adversarial scheduling instance
that confounds \lssf. For simplicity, we describe a scheduling
instance where job parameters may be irrational because of square
roots. We then show how to round the job parameters so that all are
rational.

Our bad instance consists of $n$ jobs, indexed by increasing arrival
time. We ensure that jobs~$2\ldots n$ have no slack, that is,
\begin{equation}\eqlabel{work}
 w_j=\frac{(d_j-r_j)^2}{2}\quad\quad (2\leq j\leq n)\,.
\end{equation}
Thus, in order for job~$j$ ($2\leq j\leq n$) to complete by its
deadline, job~$j$ must be executed without pause during its entire
interval.
In contrast, job~$1$ does have slack.

We arrange jobs~$1$-$3$ so that in \lssf, job~$3$ does not begin
executing until after its due date $d_3$. To do so, we assign
intervals for jobs~$1$ and~$2$ so that $r_2>r_1$, $d_2<d_1$, and
$d_1-r_1 = O(1)$.
Thus, in the \lssf schedule, job~$1$ works uninterrupted until some
point in job~$2$'s interval when job~$2$ has the largest
stretch-so-far and so begins executing. Since job~$2$ has no slack,
it finishes late, after its deadline $d_2$. We  place job~$3$ so its
release time is job~$2$'s deadline and its deadline is job~$2$'s
completion time, i.e., $r_3=d_2$ and $d_3=C_2$. In \lssf, job~$3$
does not start until its due date, $d_3$, and then works
uninterrupted until it completes for an interval stretch of
$s_3=\sqrt{2}$; see \figref{lssf1and2}(a).

We now assign jobs~$4\ldots n$ as follows; see
\figref{lssf1and2}(b). Each job $j$ has release time
\begin{equation}\eqlabel{release-time}
r_j = d_{j-1}\quad \quad (3\leq j\leq n)\,.
\end{equation}
Moreover, in \lssf we assign $d_j$ so that job~$j$ has a
stretch-so-far at time $C_{j-1}$ of
\begin{equation}\eqlabel{stretch}
\frac{C_{j-1}-r_{j}}{d_{j}-r_{j}} = s_{j-1} =
\frac{C_{j-1}-r_{j-1}}{d_{j-1}-r_{j-1}}
 \quad\quad (4\leq j\leq n)\,.
\end{equation}

\begin{figure}[t!]
\centering
{\sf (a)}\includegraphics[scale=0.8]{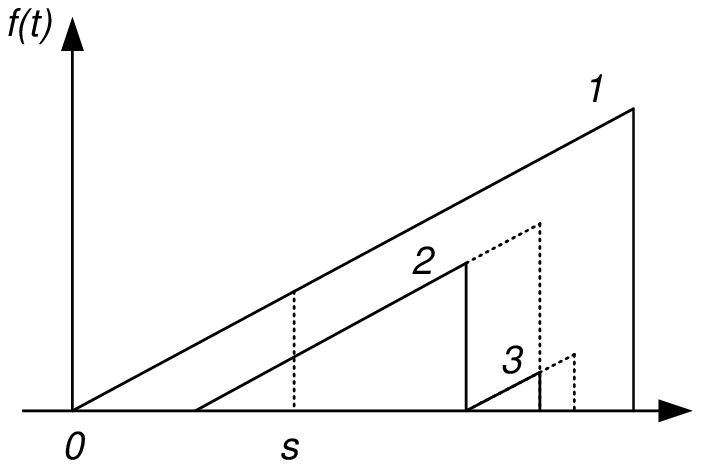}%
{\sf (b)}\includegraphics[scale=0.8]{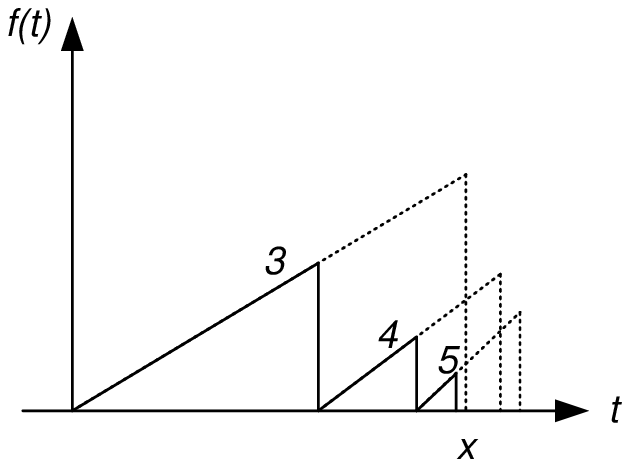}
\caption{(a) Job $2$ starts work at $s$ and completes after its due
date.  Job $3$ starts work at its due date and finishes with a
stretch of $\sqrt{2}$ (b) A stream of jobs with increasing stretch.
The stretch of job~$4$ is $\sqrt{2}$ when it starts work at time $x$
and $\sqrt{3}$ when it finishes. \figlabel{lssf1and2}}
\end{figure}

In the following we  show that in \lssf, $s_n=\Theta(\sqrt{n})$. In
contrast, in \opt, all jobs finish before their deadlines:
jobs~$2\ldots n$ run in their intervals and job~$1$ begins before
job~$2$ arrives and completes after job~$n$ completes.

We now analyze the performance of \lssf on these $n$ jobs.

\begin{theorem}\thmlabel{theorem5}
There exist feasible scheduling instances of $n$ jobs for which the
scheduling policy \lssf achieves a max-stretch of
$\Theta(\sqrt{n})$.
\end{theorem}

\begin{proof}
We analyze the performance of \lssf on the schedule instance given
above. We derive a recurrence for the stretch $s_j$ as a function of
$s_{j-1}$. Then we solve the recurrence, showing that
$s_n=\Theta(\sqrt{n})$.

Define interval $I_j=d_j-r_j$.
Recall that in \lssf, job~$j$ only begins executing at time
$C_{j-1}=r_j+s_{j-1}I_j$ because, from \eqreftwo{release-time}{stretch}, while
job~$j-1$ is in the system, its stretch-so-far is larger than that of
job~$j$'s.

We now determine the time $x_j$ that job~$j$ spends running.
By Equations~\eqreftwo{work}{stretch}, we have
 $$
 w_{j}=\frac{I_j^2}{2}=s_{j-1}I_jx_j+\frac{x^{2}_j}{2}
 \quad\quad(j\geq 4)\,.
 $$
Solving for $x_j$ and taking the positive root, we obtain
\begin{equation}\eqlabel{xj}
  x_j = -s_{j-1}I_j + I_j\sqrt{1+s_{j-1}^2}  \quad\quad(j\geq 4)\,.
\end{equation}
Thus, the stretch is
 $$
 s_j = \frac{C_{j-1}+x_j-r_{j}}{I_j}\quad\quad(j\geq 4)\,.
 $$
From \eqref{stretch}, the previous equation simplifies to
 $$
 s_j =s_{j-1}+\frac{x_j}{I_j}\quad\quad(j\geq 4)\,.
 $$
Finally, from \eqref{xj} we substitute for $x_j$, obtaining
 $$
 s_j =\sqrt{1+s_{j-1}^2} \quad\quad(j\geq 4)\,.
 $$
The solution to this
recurrence is
\begin{equation}\eqlabel{stretch-solution}
  s_j=\sqrt{j-1}\quad\quad(j\geq 3)\,,
\end{equation}
meaning that the max-stretch is $s_{n}=\sqrt{n-1}$.

We now show how big job~$1$'s interval has to be for the entire
scheduling instance to be feasible.
We make a recurrence for the interval length $I_j$. By
\eqref{stretch}, we obtain
$$
  I_j=I_{j-1}\left(\frac{C_{j-1}-r_{j}}{C_{j-1}-r_{j-1}}\right)
=I_{j-1}\left(1+\frac{r_{j-1}-
r_{j}}{C_{j-1}-r_{j-1}}\right)\quad\quad(j\geq 4)\,.
$$
Finally, by Equations~\eqreftwo{release-time}{stretch-solution}, we
obtain
 \begin{eqnarray*}
  I_j=I_{j-1}\left(1+\frac{1}{s_{j-1}}\right)
  =I_{j-1} \left(1-\frac{1}{  \sqrt{j-2}}\right)\quad\quad(j\geq 4)\,.
 \end{eqnarray*}
Therefore, assuming w.l.o.g.\ that $I_3=1$, an upper bound on $I_j$
is
\begin{eqnarray*}
I_j=\prod_{i=2}^{j-2}\left(1-\frac{1}{\sqrt{i}}\right)\leq
e^{-\sqrt{j-3}} \quad\quad(j\geq 3)\,.
\end{eqnarray*}
The sum of all intervals lengths is
\begin{eqnarray*}
\sum_{j=3}^{n} I_j=\sum_{j=3}^{n}e^{-\sqrt{j-3}}=O(1)
 \quad\quad(j\geq 3)\,.
\end{eqnarray*}
Consequently, it suffices to set $I_1=O(1)$ and $w_1=O(1)$  to
obtain a feasible schedule.\qed
\end{proof}

This particular example has job parameters that may be irrational
(because of square roots). In fact, we can come up with another
scheduling instance so that the input size (number of bits necessary
to describe the scheduling instance) is polynomial in $n$. The idea
is to round job parameters so that all are rational. We round the
interval length $d_j-r_j$ of job $j$ up to a rational number and
round the work $w_j$ down to a rational number. We make both $r_j$
and $d_j$ rational and retain Equation~\eqref{release-time}. We make
the equality in Equations~\eqreftwo{work}{stretch} only approximate,
that is, for arbitrarily small nonnegative $\epsilon_j$ and
$\epsilon_j'$,
\begin{eqnarray*}
 w_j+\epsilon_j
 ={(d_j-r_j)^2}/{2}\quad &(2\leq j\leq n) \nonumber\\
 \frac{C_{j-1}-r_{j}}{d_{j}-r_{j}} + \epsilon_j'
 = s_{j-1}  \quad\quad &(4\leq j\leq n)\,.\nonumber
\end{eqnarray*}
The analysis for \thmref{theorem5} carries over.

\subsection*{\boldmath $\Theta(1)$-Competitive Online Algorithm for Max-Stretch}

We now exhibit the strategy \thrashing,
which bounds the interval stretch of each job by $4$. The \thrashing strategy
models the extreme case of a procrastinator who does not
work on any job until it has already passed its due date.  More formally, in
the this strategy no job is executed until it has a stretch of at
least $2$. Among all such jobs, the procrastinator executes the job
that arrived latest.

Before proceeding, we explain our choice of terminology. An
operating system is said to `thrash' when it begins running
inefficiently because it spends too much time context switching.
`Thrashing' is now also commonly used among computer scientists to
describe their own behavior when they have too many jobs to finish.
The name is applied here because the procrastinator appears to be
thrashing. Each time a more recent job has too large an interval
stretch, the procrastinator abandons the current job and executes
the more recent job.

We begin by proving the following simple lemma:

\begin{lemma}
Consider a feasible set of jobs $1,\ldots,m$ and consider times $r$
and $d$, where all $r_j\geq r$ and $d_j\leq d$.  Let \delay be any
scheduling policy that only schedules work from jobs having stretch
at least $\alpha$, where $\alpha\geq1$.  The total amount of time
required to run all jobs using \delay is at most $(d-r)/\alpha$.
\lemlabel{delayedwork}
\end{lemma}

\begin{proof}
Because the set of jobs is feasible, there is some way to schedule
each job within its interval and the total time spent working is at
most $d-r$.  Now consider running \delay.  For any given job $j$,
the slowest that $j$ runs in \delay is at least $\alpha$ times
faster than $j$ runs in the feasible schedule.  The lemma follows
immediately.\qed
\end{proof}

\begin{theorem}\thmlabel{2-lifo}
For any feasible set of jobs, \thrashing bounds the interval stretch of
every job by~$4$.
\end{theorem}

\begin{proof}
The proof is by contradiction. Define the \emph{extended due date}
$\extendedj$ of job $j$ to be the time that $j$ must complete by to
guarantee an interval stretch of~$4$, that is,
$\extendedj=4(d_j-r_j)+r_j$.  Consider some job~$j$ that does not
meet its extended due date.  For simplicity and without loss of
generality, we normalize time so that $r_j=0$ and $d_j=1$. Job~$j$
cannot begin until time~$2$ and by assumption completes at some time
$f>4$.

By \lemref{delayedwork}, the total amount of time spent working on
all jobs (including $j$) whose intervals are entirely contained
within $[0,4]$ is at most $4/2=2$ units of time.  Moreover, there
can be no gaps in the schedule during the interval $[2,f]$ because
otherwise $j$ would work during the gaps and finish earlier than
time~$f$. Finally, by the definition of \thrashing, there can be no
work scheduled during~$[2,f]$ on jobs having release dates
before~$0$ because $j$ has higher priority. Thus, $f$ cannot be
greater than $4$ and we obtain a contradiction.\qed
\end{proof}

It may, of course, be unrealistically optimistic to give the online
procrastinator the power to run arbitrarily fast.  However, it
follows from \thmref{2-lifo} that \thrashing never runs any job $j$
faster than $4f_j(d_j)$.  In fact, the proof of \thmref{2-lifo}
indicates that we can reduce this upper bound still further to
$2f_j(d_j)$ without increasing the max-stretch; we need only modify
the speed functions so that the maximum job speed for job $j$ is
limited to $2f_j(d_j)$.

\section{Conclusions}

The first sentence of the conclusion, which summarizes the paper, is
being written just a few hours before the deadline. As we were
writing this paper, we were struck by the wealth of open problems in
this area. For example, what is the right way to resolve the
computational and numerical issues associated with linear and other
speed functions? The scheduling problem (even in the offline case)
becomes even more complex with speed functions that may be nonzero
at jobs' release times. (This is because \lrtb fails, and the
optimal schedule seems to depend on the workload as well as on the
slopes of the speed functions.) For our online algorithm we did not
try to optimize the constant in the online competitive ratio fully;
what is the smallest that we can make this constant, especially
where the speed functions are sublinear?

We have also considered piecewise-constant speed functions and have
linear programming solutions for several variants of the problem.
The LP has constraints for each time interval $[t_1,t_2]$ in which
the execution speeds of all jobs are constant. (Specifically, within
$(t_1,t_2)$ there are no job release times or deadlines, and for
each job $j$ the function of $f_j(t)$ is constant when
$t\in[t_1,t_2]$.) There are many metrics we can optimize. For
example, we can minimize or maximize the total amount of time
working.  Alternatively, we can introduce a notion of stress for the
procrastinator and find the least stressful schedule.

Finally, what about other metrics, especially in models where some
jobs may be left unexecuted? What about settings where job streams
are executed on parallel processors?

It is now several hours later, just minutes before the deadline. We
were searching for the ideal way to end the paper and circumstances
have unfortunately provided the answer. A campus-wide power failure
at Stony Brook has cut two hours from our last-minute working time
and highlights the difficulties of online scheduling for
procrastinators.

\subsection*{Acknowledgments}

{\small We are grateful to Esther Arkin, Nikhil Bansal, and Joseph
Mitchell for many helpful discussions. We thank Nikhil Bansal for
the LP solution for piecewise constant speed functions.}

{\footnotesize
\bibliographystyle{abbrv}
\bibliography{procrastinate}
}

\end{document}